# Amphoteric behavior of Hydrogen in Bimetallic Molecular like Hydrides

S Kiruthika[1, 2], R Varunaa[1, 2] and P Ravindran[1, 2, 3, a)]

[1] *Department of Physics, Central University of Tamil Nadu, Thiruvarur-610005*
[2] *Simulation Centre for Atomic and Nanoscale MATerials, Central University of Tamil Nadu, Thiruvarur-610005*
[3] *Department of Chemistry and Centre for Materials Science and Nanotechnology, University of Oslo, P.O. Box: 10333, Blindern, N 1035, Oslo, Norway.*
[a)] *Corresponding author: raviphy@cutn.ac.in*

**Abstract.** Generally hydrogen will adopt the +1, 0 or -1 oxidation state in solids depending upon the chemical environment it occupy. Typically, there are some exceptional cases in which hydrogen exhibits both anionic and cationic behavior in the same structural frame works. In this study we briefly explore an amphoteric behavior of hydrogen present in ammine bimetallic borohydrides with the chemical formula $M_1M_2(BH_4)_3(NH_3)_2$ ($M_1$= Na; $M_2$ = Zn, Mg) using the state-of-the-art density functional calculations. In order to establish the amphoteric behavior of hydrogen in $M_1M_2(BH_4)_3(NH_3)_2$, we have made detailed chemical bonding analyses using partial density of states, charge density, electron localization function, and Born effective charge. From these analyses we found that the hydrogen closer to boron is in negative oxidation state whereas the hydrogen closer to nitrogen is in positive oxidation state. The establishment of the present of hydrogen with amphoteric behavior in solids has large implication on hydrogen storage application.

## INTRODUCTION

Hydrogen is an attractive and pollution-free energy carrier. Hence it is considered as a sustainable energy source. Safe storage of hydrogen at high volumetric and gravimetric density with appropriate dehydrogenation temperature is still a challenging task towards the hydrogen economy. It is evident that solid state hydrogen storage methods are the most promising way to store hydrogen at the higher capacity. The metal ammine and metal borohydrides are considered as potential hydrogen storage materials due to their high hydrogen storage capacity. Though the metal borohydrides are thermodynamically stable, their decomposition temperature is very high for the release of hydrogen which limits their use in wide range of practical applications. Recently, it has been reported that the thermodynamic properties of ammine metal borohydrides can be tuned by using different metal cations or multiple cations substitutions[1]. Interestingly, the hydrogen present in these materials exhibits both +1 and -1 oxidation state so called amphoteric behavior.

Ravindran *et al.,*[2–4] have already reported that it is possible to violate the "2-Å rule" for H-H separation in hydrides and this open up the possibility to design new hydrogen storage materials with high volume efficiency. Based on this, we are trying to design efficient hydrogen storage materials with H-H separation less than 2 Å. In this mission we are looking for systems were hydrogen is in both + and – oxidation state so that we can keep the hydrogen ions very short H-H distance in such systems due to attractive Coulombic interaction between these oppositely charged ions. So, it is interesting to find compounds with hydrogen exhibiting amphoteric nature. We believe that this idea can help in designing hydrogen storage materials with high volume density. In this work, we have interpreted the amphoteric behavior of hydrogen in bimetallic ammine borohydrides (BMAB) with the help of various tools from first principle techniques such as density of states, charge density plot, electron localization function, Born effective charge, etc.

## COMPUTATIONAL DETAILS

The total energy calculations have been performed using density functional theory (DFT) with generalized gradient–approximation (GGA), as implemented in the Vienna *ab-initio* simulations package (VASP)[5]. All the calculation were carried out with a 300 eV plane wave cutoff. The structural optimizations were performed using force as well as stress minimization methods. For the structural optimization, the **k**-points were generated using the Monkhorst-Pack method with the grid of 6×4×2. For the density of state calculations we have used the tetrahedron method with higher **k**-point density.

## RESULT AND DISCUSSIONS

The particular compounds studied here are the newly predicted $M_1M_2(BH_4)_3(NH_3)_2$ where $M_1$= Na; $M_2$ = Zn, Mg[6] composition. Fig.1 (a) shows the optimized orthorhombic crystal structure of $M_1M_2(BH_4)_3(NH_3)_2$ with the space group of P21nb [1,6]. We have used $NaZn(BH_4)_3(NH_3)_2$ structural data to optimize the crystal structure of $NaMg(BH_4)_3(NH_3)_2$ by replacing Zn with Mg. In these BMABs, the divalent cation Zn/Mg is tetrahedrally coordinated with $BH_4$ and $NH_3$ structural sub-units. We have denoted the H closer to B as $H^{-1}$ and that closer to N as $H^{+1}$ since these two hydrogen atoms are in different oxidation state as follows.

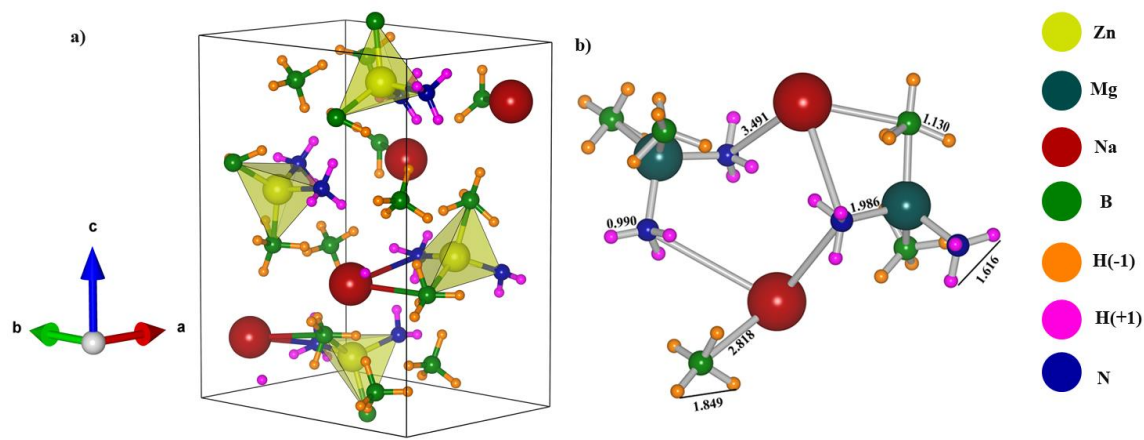

**FIGURE 1.** Ground state crystal structure of (a) $NaZn(BH_4)_3(NH_3)_2$ and (b) the bond distance of $NaMg(BH_4)_3(NH_3)_2$

Fig. 2(a) and 2(b) show the total and partial DOS for $NaZn(BH_4)_3(NH_3)_2$ and $NaMg(BH_4)_3(NH_3)_2$, respectively.

The total density of states (DOS) indicate that these BMAB systems are insulators with wide band gap of around 5 eV. From the partial DOS of $NaZn(BH_4)_3(NH_3)_2$ (see Fig. 2(a)), it can be seen that the valence band (VB) is mainly originate from Zn-*d* states and minor contributions from Zn-*s, p*, B-*p*, $H^{-1}$-*s*, and N-*p* states. Similarly, the partial DOS of $NaMg(BH_4)_3(NH_3)_2$ (see Fig.2 (b)) reveals that the valence band is mainly composed of Mg-*s, p*, B-*p*, N-*p*, and $H^{-1}$-*s* states. The negligible DOS present at the Na-site indicate that it is ionically bonded with the neighboring $BH_4$ and $NH_3$ structural sub-units. It is to be noted that B-*2p* and $H^{-1}$-*s* states are energetically degenerate over the energy range of -0.5 eV to -2.5 eV whereas in the lowest energy region of VB (around 6 eV), larger value of DOS from B-*2s* states are present compared with that from $H^{-1}$-*s* states. This indicates a mixed bonding behavior (iono-covalent) is present in the $BH_4$ structural sub-unit. In contrast to $BH_4$ unit, the $NH_3$ structural sub-unit has relatively higher ionic bonding character since the DOS of N-*2p* states are well separated to that of $H^{+1}$-*s* states in the VB apart from very small contribution from $H^{+1}$-*1s* states present at the higher energy region of VB. Also, it is interesting to note that negligible amount of DOS are present at the $H^{+1}$ site whereas noticeable amount of DOS are present at the $H^{-1}$ site. Form this illustration, it is clearly evident that amphoteric hydrogens are present in the BMAB systems.

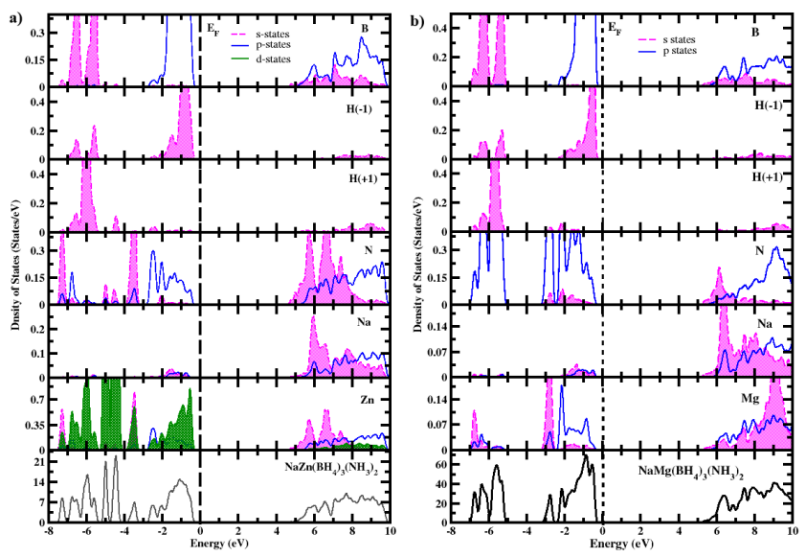

**FIGURE 2.** Total and partial DOS for (a) $NaZn(BH_4)_3(NH_3)_2$ and (b) $NaMg(BH_4)_3(NH_3)_2$, respectively.

In order to substantiate the amphoteric behavior of hydrogen and also the nature of chemical bonding between the hydrogen with their neighbors in BMAB systems, we have plotted charge density distribution and electron localization function (ELF) for $NaZn(BH_4)_3(NH_3)_2$ and $(NaMg(BH_4)_3(NH_3)_2$ in Fig. 3(a) and 3(b), respectively. The spherical charge distribution around the Zn atom in Fig 3(a) and negligible charges at Na site reveals that there is an ionic bonding present between the metal ions Zn/Na and the molecule-like $BH_4$ and $NH_3$. The relatively long B-H distance over N-H distance along with B has relatively small (with respect to N) spherical distributed charge with small charge between B and $H^{-1}$ in $BH_4$ molecular sub-unit, indicating noticeable ionic bonding character. Due to H in -1 charge state near B and +1 charge state near N, the calculated charge density clearly show wide spatial distribution at the H site when it is closer to B than that closer to N clearly demonstrate its amphoteric behavior.

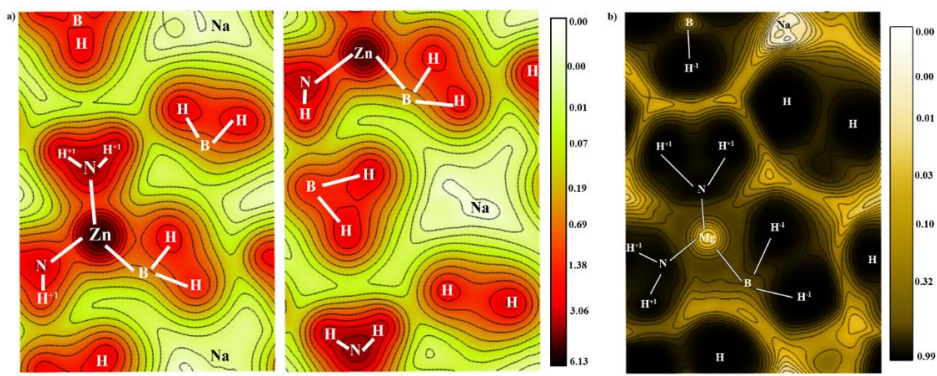

**FIGURE 3.** (a) Charge density distribution for $NaZn(BH_4)_3(NH_3)_2$ and (b) electron localization function for $NaMg(BH_4)_3(NH_3)_2$ obtained from DFT calculations.

Figure 3(b) shows the ELF plot for $NaMg(BH_4)_3(NH_3)_2$ in which within $BH_4$ molecular sub-units higher value of ELF closer to one present between H and also between B and H indicate the noticeable covalent bonding. Further, localized charges are present in the H site in $BH_4$-unit indicate that H is the -1 oxidation state. In the case of $NH_3$-sub-unit the ELF is more polarized towards N with a value of one indicating that the H in the $NH_3$-molecular unit are in +1 oxidation state. The ELF at the metal site (Mg/Zn) clearly shows that there is dissipation of electron charge around it indicates the ionic interaction between the metal (Mg/Zn) atom and the $BH_4/NH_3$ molecular unit. Therefore, from the charge density distribution and ELF plot, it is concluded that there is an ionic bonding between metal ions and the $BH_4/NH_3$ molecular unit and a mixed iono-covalent bonding between B and H. The bonding interaction between N and H is dominantly ionic with small character.

**TABLE 1.** Born effective charge analysis for NaZn(BH$_4$)$_3$ (NH$_3$)$_2$ and NaMg(BH$_4$)$_3$ (NH$_3$)$_2$

| Compound Name | Atom | Z*(e) | | | | | | | | |
|---|---|---|---|---|---|---|---|---|---|---|
| | | xx | yy | zz | xy | yz | zx | xz | zy | yx |
| NaZn(BH$_4$)$_3$(NH$_3$)$_2$ | Na | 1.070 | 1.092 | 1.143 | -0.125 | 0.071 | 0.136 | 0.152 | 0.022 | -0.106 |
| | Zn | 1.751 | 1.602 | 1.785 | -0.090 | -0.198 | 0.102 | 0.162 | -0.165 | -0.037 |
| | B | 0.155 | 0.201 | 0.200 | -0.003 | -0.033 | 0.108 | 0.016 | 0.242 | -0.007 |
| | H$^{-1}$ | -0.179 | -0.529 | -0.326 | -0.011 | 0.206 | 0.018 | 0.016 | 0.242 | -0.007 |
| | H$^{+1}$ | 0.461 | 0.264 | 0.210 | 0.070 | -0.008 | -0.113 | -0.103 | -0.001 | 0.056 |
| | N | -0.649 | -0.932 | -0.782 | -0.009 | -0.123 | 0.001 | -0.036 | -0.085 | 0.066 |
| NaMg(BH$_4$)$_3$(NH$_3$)$_2$ | Na | 1.052 | 1.055 | 1.110 | -0.118 | -0.069 | -0.114 | -0.123 | -0.046 | -0.115 |
| | Mg | 1.732 | 1.553 | 1.851 | -0.020 | -0.169 | 0.162 | 0.178 | -0.145 | 0.001 |
| | B | 0.079 | 0.174 | 0.134 | 0.001 | -0.026 | 0.083 | 0.012 | -0.005 | 0.006 |
| | H$^{-1}$ | -0.178 | -0.508 | -0.302 | -0.002 | 0.193 | 0.020 | 0.018 | 0.220 | -0.011 |
| | H$^{+1}$ | 0.432 | 0.279 | 0.222 | 0.036 | 0.015 | -0.086 | -0.094 | 0.019 | 0.027 |
| | N | -0.681 | -0.810 | -0.802 | -0.023 | -0.086 | -0.006 | -0.016 | -0.038 | 0.040 |

We have also calculated the Born effective charge (BEC) to reveal chemical bonding between constituents and amphoteric behavior of hydrogen. The calculated Born effective charge tensor (Z*) of BMAB systems are tabulated in Table 1. The diagonal BEC tensors of Na are equal to the nominal ionic charge (+1) showing that Na is ionically bonded to the molecular subunit. The other metal ions i.e. Zn and Mg have smaller BEC value than nominal charge of 2+. Similarly, the BEC value of B, N, H$^{-1}$ and H$^{+1}$ are smaller than the nominal charges (+3, -3, +1 and -1). The diagonal components of BEC tensor are not equal, and the off-diagonal tensor have finite value showing the anisotropy in charge distribution which is one of the properties of the covalent bonds. These observations conclude the presence of mixed i.e. iono-covalent type of bonding in these systems.

## CONCLUSION

We have performed first principle calculations for NaMg/Zn(BH$_4$)$_3$(NH$_3$)$_2$. Our results show that these systems are insulators with band gap of around 5 eV. We have shown that the metal ions (Na, Zn and Mg) have pure ionic bonding with the molecular subunits i.e. BH$_4$ and NH$_3$ whereas, in the subunits the bonding between the N-H and B-H have iono-covalent nature. The hydrogen closer to the boron has the negative oxidation state and that closer to the nitrogen has positive oxidation state. Therefore, our detailed analysis revealed the amphoteric nature of hydrogen in the above studied compounds.

## ACKNOWLEDGMENTS

The authors are grateful to the Department of Science and Technology, India for the funding support via Grant No. SR/NM/NS-1123/2013 and the Research Council of Norway for providing computer time (under the project number NN2875k) at the Norwegian supercomputer facility. This research was also supported by the Indo-Norwegian Cooperative Program (INCP) via Grant No. F. No. 58-12/2014(IC) and Research Council of Norway, Grant No. 221905(FRIPRO).